\documentclass[twocolumn,showpacs,preprintnumbers,amsmath,amssymb]{revtex4}
\usepackage{graphicx}
\usepackage{color}
\usepackage{dcolumn}
\usepackage{bm}

\begin{document}
\title{Bulk induced phase transition in driven diffusive systems}
\author{Yu-Qing Wang$^{1}$}
\author{Rui Jiang$^{1}$}\email{rjiang@ustc.edu.cn}
\author{Anatoly B Kolomeisky$^{2}$}
\author{Mao-Bin Hu$^{1}$}

\affiliation{$^{1}$ State Key Laboratory of Fire Science and
School of Engineering Science, University of Science and
Technology of China, Hefei 230026, China }

\affiliation{$^{2}$ Department of Chemistry, Rice University,
Houston, TX 77005-1892, USA}
\date{\today}

\begin{abstract}
This Letter studies a weakly and asymmetrically coupled three-lane
driven diffusive system. A non-monotonically changing density
profile in the middle lane has been observed. When the extreme
value of the density profile reaches $\rho=0.5$, a bulk induced
phase transition occurs which exhibits a shock and a continuously
and smoothly decreasing density profile which crosses $\rho=0.5$
upstream or downstream of the shock. The existence of double
shocks has also been observed. A mean-field approach has been used
to interpret the numerical results obtained by Monte Carlo
simulations. The current minimization principle has excluded the
occurrence of two or more bulk induced shocks in the general case
of nonzero lane changing rates.

\end{abstract}


\maketitle

{\it Introduction.} Driven diffusive system is a rewarding
research topic in recent decades, which exhibits non-vanishing
current even in the steady state and has served as fruitful
testing grounds for fundamental research in non-equilibrium
physics [1-5]. The driven diffusive systems exhibit many
surprising or counterintuitive features, given our experiences
with equilibrium systems, e.g., spontaneous symmetry breaking,
phase separation, etc. [6-10].

The boundary induced phase transition is another non-equilibrium
phenomenon [11-14]. For the case of vanishing right boundary
density, Krug postulated a rather general maximal-current
principle that the system tries to maximize its stationary current
[11]. The maximal-current principle has later been generalized to
the extremal current principle [12,13]
\begin{equation}
j=\left\{
\begin{aligned}
\max~~j(\rho)~~~\text{for}~\rho_{-}>\rho_{+}~~ \\
\min~~j(\rho)~~~\text{for}~\rho_{-}<\rho_{+}~~\\
\end{aligned}
\right.
\end{equation}
Here $\rho_-$ ($\rho_+$) is the constant effective density of the
left (right) reservoir from which particles are flowing into (out
of) the system. The microscopic details of the system only
determine the functional form of the current $j(\rho)$ and the
effective boundary densities.

Motivated by facts such as the unidirectional motion of many motor
proteins along cytoskeletal filaments, in which motors advance
along the filament while attachment and detachment of motors
between the cytoplasm and the filament occur, the constraint of
the conserved dynamics in the bulk in the driven diffusive system
has been relaxed to consider random particle attachments and
detachments in the bulk [15]. The resulting dynamics leads to a
phase coexistence of low and high density regions separated by a
shock.

\begin{figure}[!h]
\centering
\includegraphics[width=0.45\textwidth]{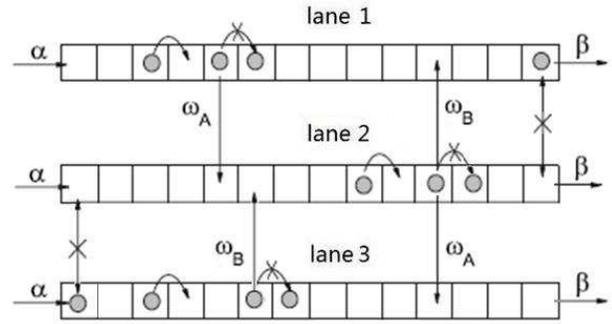}
\caption{Sketch of the three-lane TASEP. The arrow shows allowed
hopping and the cross shows prohibited hopping. }
\end{figure}

The driven diffusive systems have also been extended to include
the possibility of transport on multiple parallel lanes to
describe phenomena such as the extraction of membrane tubes by
molecular motors, macroscopic clustering phenomena, car traffic
and so on (see e.g. Refs.[16-29] and references therein), in which
the boundary induced phase transitions still could be observed.

{\it Model and mean field equations.} This Letters studies a
weakly and asymmetrically coupled three-lane totally asymmetric
simple exclusion process (TASEP), and reports a bulk induced phase
transition. Our model is defined in a three-lane lattice of
$N\times 3$ sites, where $N$ is the length of a lane (see Fig.1).
For each time step, a site is chosen at random. A particle at site
$i$ and lane $j$ can jump with rate 1 to site $i+1$ if it is
unoccupied. Otherwise, if site $i+1$ is occupied, the particle
jumps to site $i$ on lane $j+1$ with rate $\omega_A$ and to site
$i$ on lane $j-1$ with rate $\omega_B$ ($\omega_B\ne \omega_A$),
provided the target site is unoccupied. Obviously, $\omega_B=0$
for lane 1 and $\omega_A=0$ for lane 3. The asymmetric coupling
makes the the middle lane different from two other lanes, and bulk
induced phase transition occurs on this lane. At the boundaries, a
particle enters from the left boundary with rate $\alpha$, and is
removed from the right boundary with rate $\beta$. As in previous
studies, a weakly coupling is considered in which $\omega_A
N=\Omega_A$ and $\omega_B N=\Omega_B$ are kept constant.

The hydrodynamic mean field equations of the system could be
written as

\begin{equation}
\frac{\partial \rho_1}{\partial t}=-(1-2\rho_1)\frac{\partial
\rho_1}{\partial
x}-\Omega_A\rho_1^2(1-\rho_2)+\Omega_B\rho_2^2(1-\rho_1)
\end{equation}
\begin{equation}\begin{array}{lll}
\frac{\partial \rho_2}{\partial t} & = &-(1-2\rho_2)\frac{\partial
\rho_2}{\partial x}
-\Omega_A\rho_2^2(1-\rho_{3})+\Omega_B\rho_{3}^2(1-\rho_2)\\&&
-\Omega_B\rho_2^2(1-\rho_{1})+\Omega_A\rho_{1}^2(1-\rho_2)
\end{array}
\end{equation}
\begin{equation}
\frac{\partial \rho_3}{\partial
t}=-(1-2\rho_3)\frac{\partial\rho_3}{\partial
x}-\Omega_B\rho_3^2(1-\rho_{2})+\Omega_A\rho_{2}^2(1-\rho_3)
\end{equation}

In the steady state, one has
\begin{equation}
(1-2\rho_1)\frac{d \rho_1}{d
x}=-\Omega_A\rho_1^2(1-\rho_2)+\Omega_B\rho_2^2(1-\rho_1)
\end{equation}
\begin{equation}\begin{array}{lll}
(1-2\rho_2)\frac{d \rho_2}{d x} & = &
-\Omega_A\rho_2^2(1-\rho_{3})+\Omega_B\rho_{3}^2(1-\rho_2)\\&&
-\Omega_B\rho_2^2(1-\rho_{1})+\Omega_A\rho_{1}^2(1-\rho_2)
\end{array}
\end{equation}
\begin{equation}
(1-2\rho_3)\frac{d \rho_3}{d
x}=-\Omega_B\rho_3^2(1-\rho_{2})+\Omega_A\rho_{2}^2(1-\rho_3)
\end{equation}
Here $\rho_i$ denotes the density on lane $i$. Via rescaling the
total length to unity, the boundary conditions
$\rho_i(0)=\min(\alpha,1/2)$ and $\rho_i(1)=\max(1-\beta,1/2)$
should be imposed properly, depending on the specific state of the
system. Via numerical solving Eqs.(5)-(7), we can obtain the
density profile of the three lanes.

{\it Results.} Figure 2 shows a typical phase diagram of a
three-lane system, in which an asymmetrical set of parameters
$\Omega_A=10$ and $\Omega_B=0.1$ is adopted. In the phase diagram,
$XYZ$ stands for the state of the three lanes.  $L$ $(H)$ means
that the corresponding lane $i$ is in low (high) density phase
$\rho_i<0.5 (\rho_i>0.5)$, $S$ means that there is a shock on the
lane (shocks $S_1$ and $S_2$ denote bulk induced shocks as
discussed below), $D$ means that there are two shocks on the lane,
$C$ means that the density profile is a Continuously and Smoothly
Decreasing one and Crosses $\rho=0.5$ (CSDC). The transitions
among $LLL$, $LLS$, $LSS$, $LLH$, $LSH$, $LHH$, $SSH$, $SHH$,
$HHH$ can be easily understood as boundary induced phase
transitions, see e.g., Refs.[19,20]. For instance, when
$\rho_3(1)$ increases to $\beta$ in the $LLL$ phase, a shock is
induced from the right boundary on lane 3 and thus the $LLL$ phase
transits into $LLS$ phase. Fig.3 shows the typical density
profiles of these states.

\begin{figure}[!h]
\centering
\includegraphics[width=0.5\textwidth]{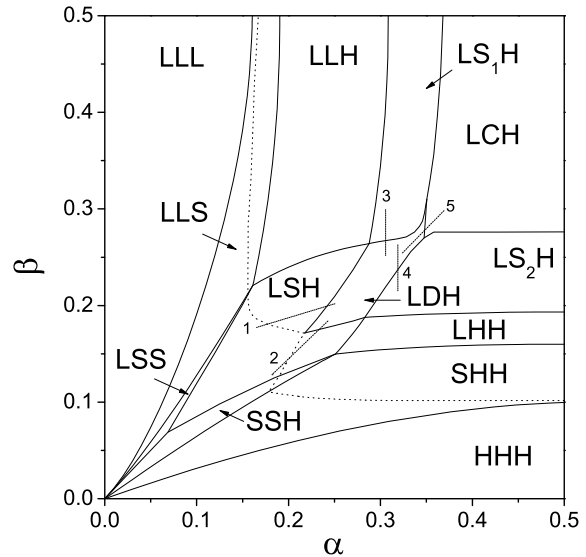}
\caption{Phase diagram of the three lane TASEP obtained from mean
field calculation. The parameters $\Omega_A=10$ and
$\Omega_B=0.1$. Details about the dotted lines and the dashed
lines 1-5 are in the text.}
\end{figure}

\begin{figure}[!h]
\centering
\includegraphics[width=0.16\textwidth]{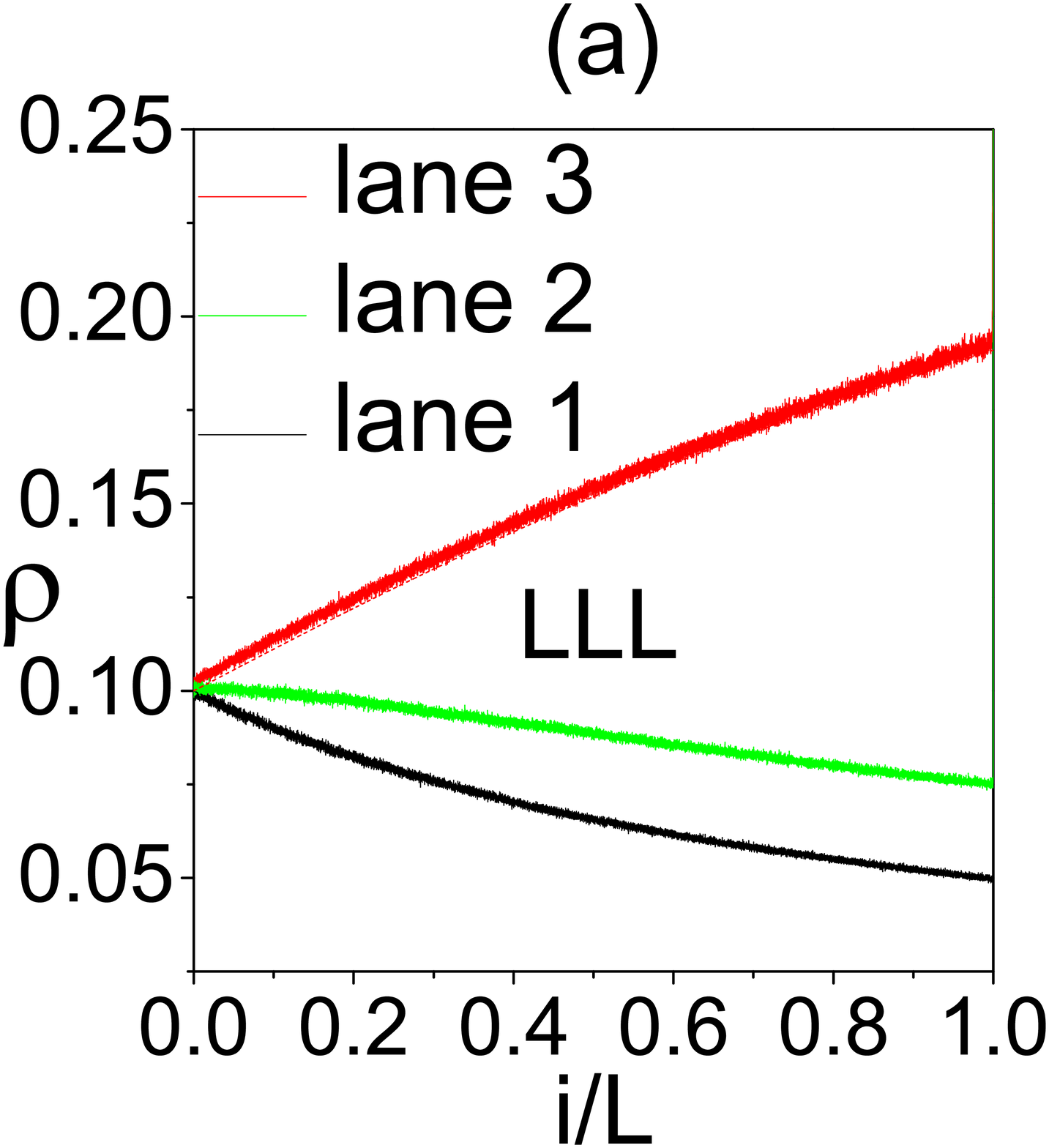}\includegraphics[width=0.16\textwidth]{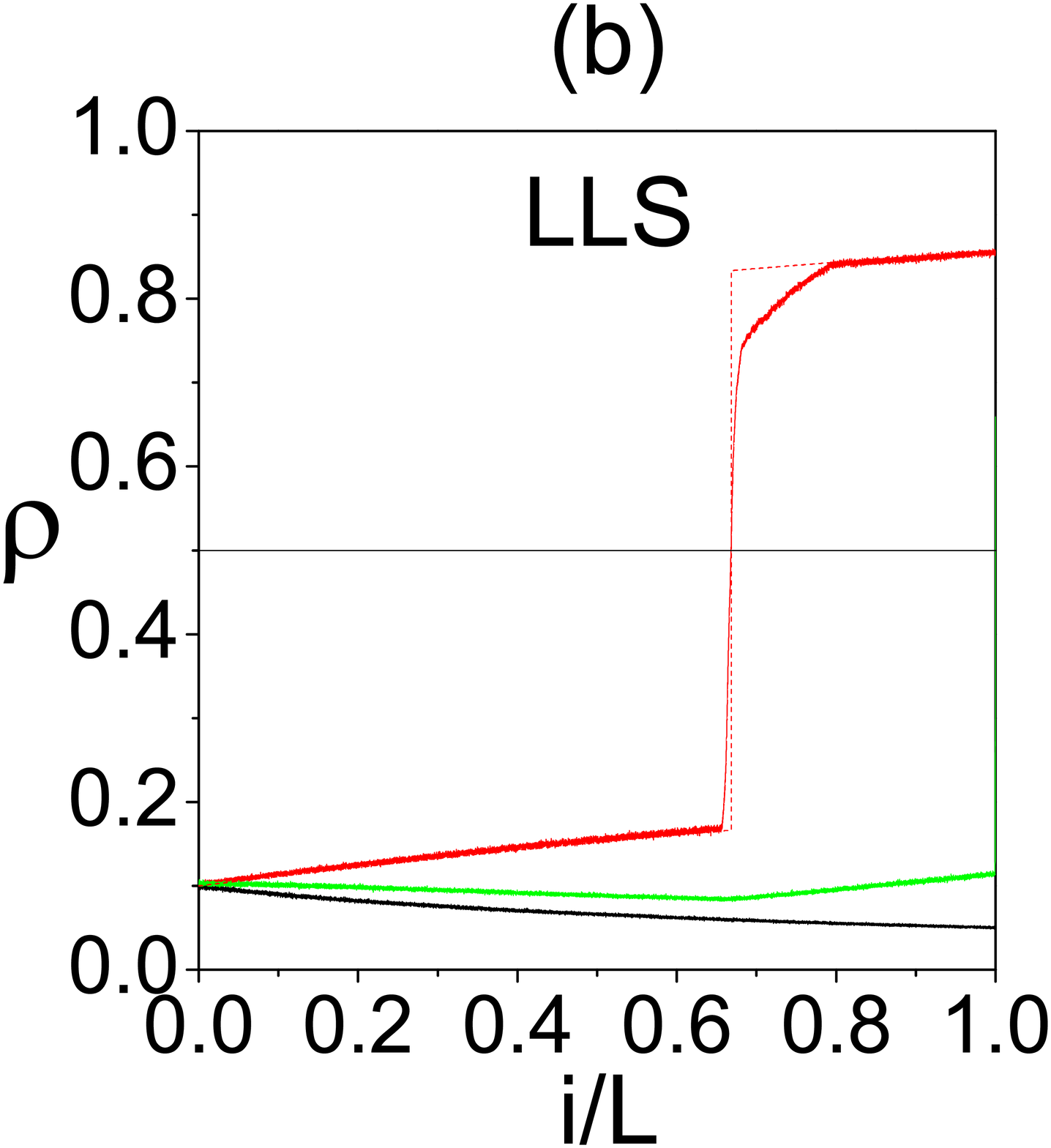}\includegraphics[width=0.16\textwidth]{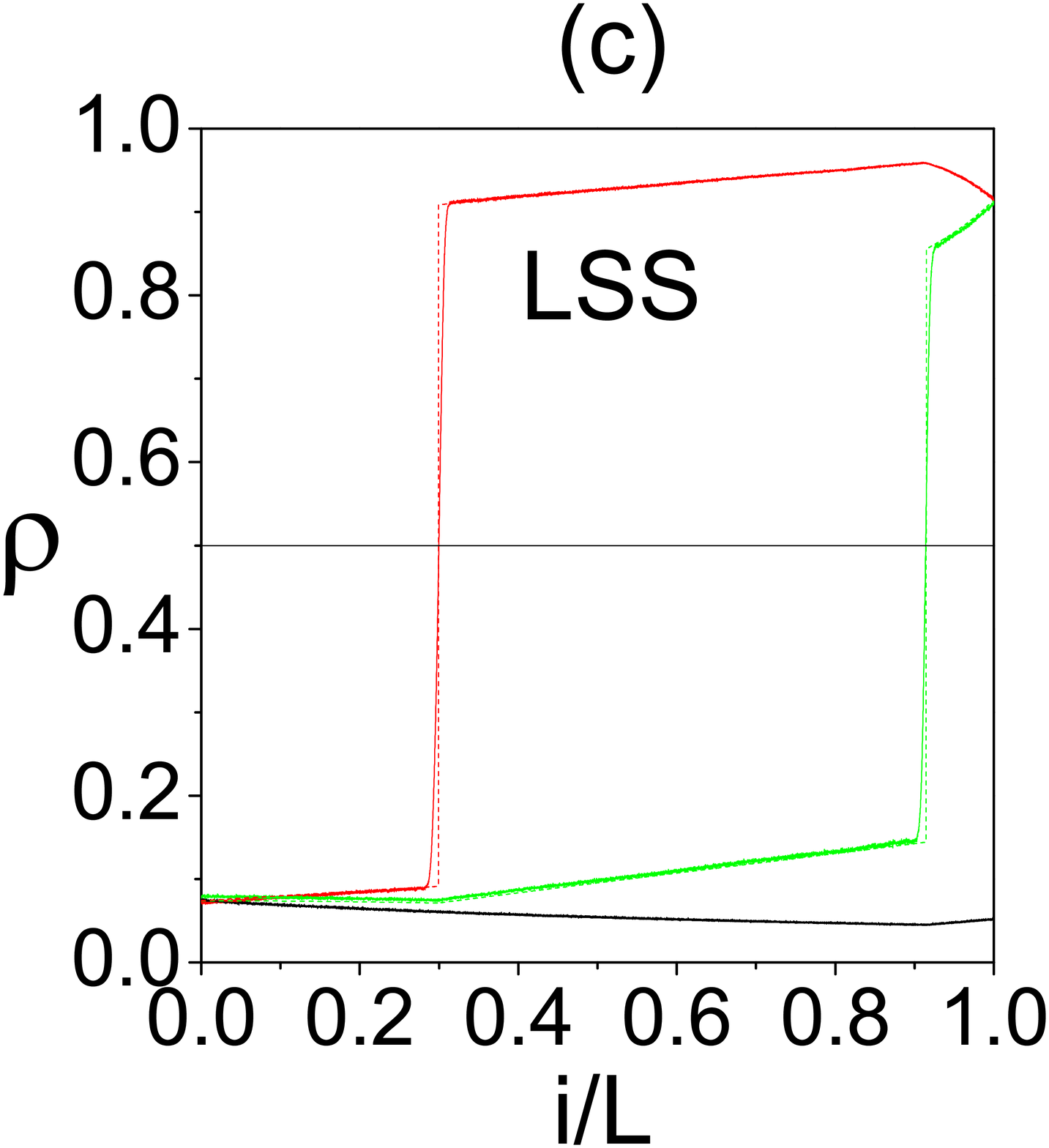}
\includegraphics[width=0.16\textwidth]{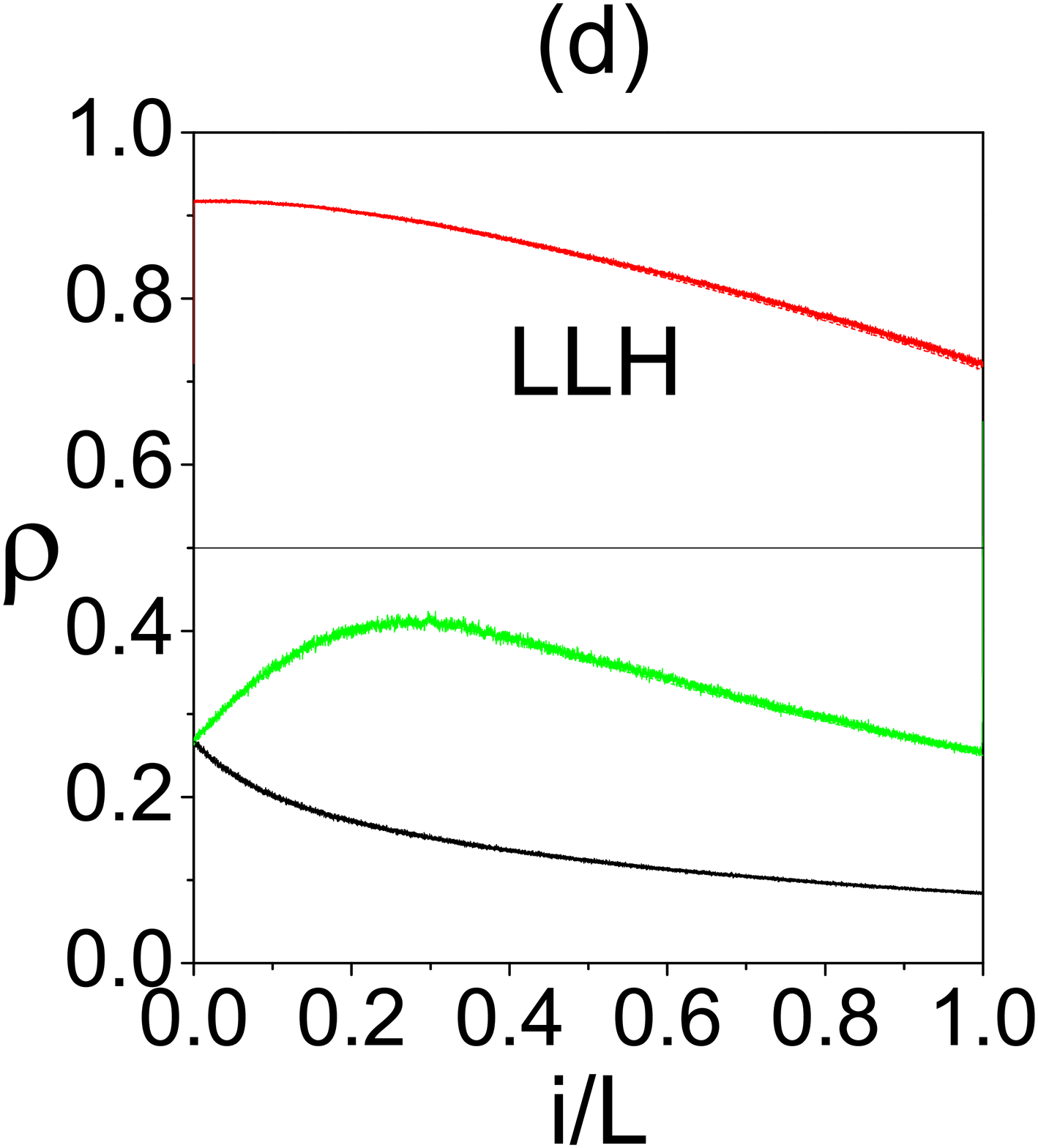}\includegraphics[width=0.16\textwidth]{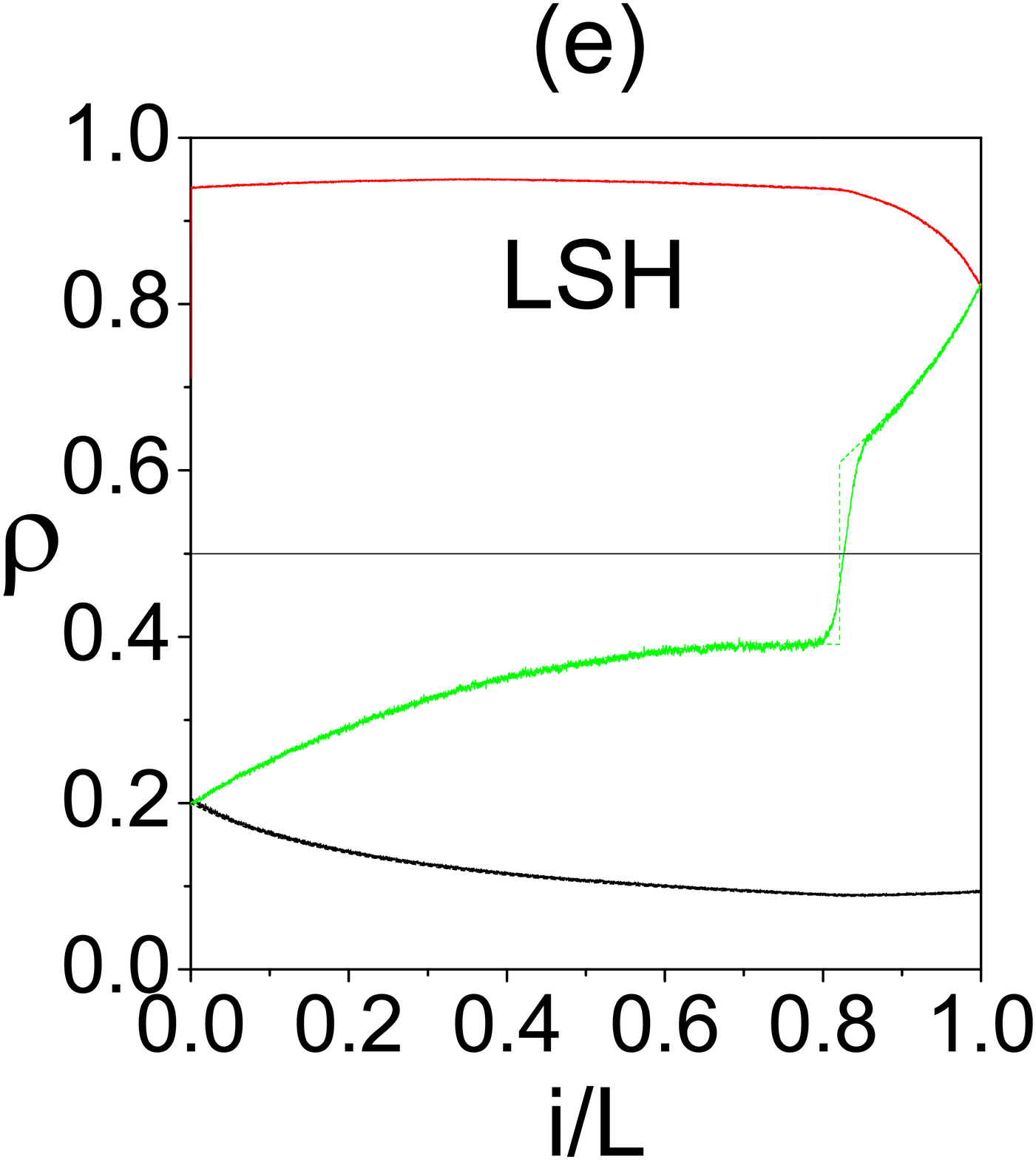}\includegraphics[width=0.16\textwidth]{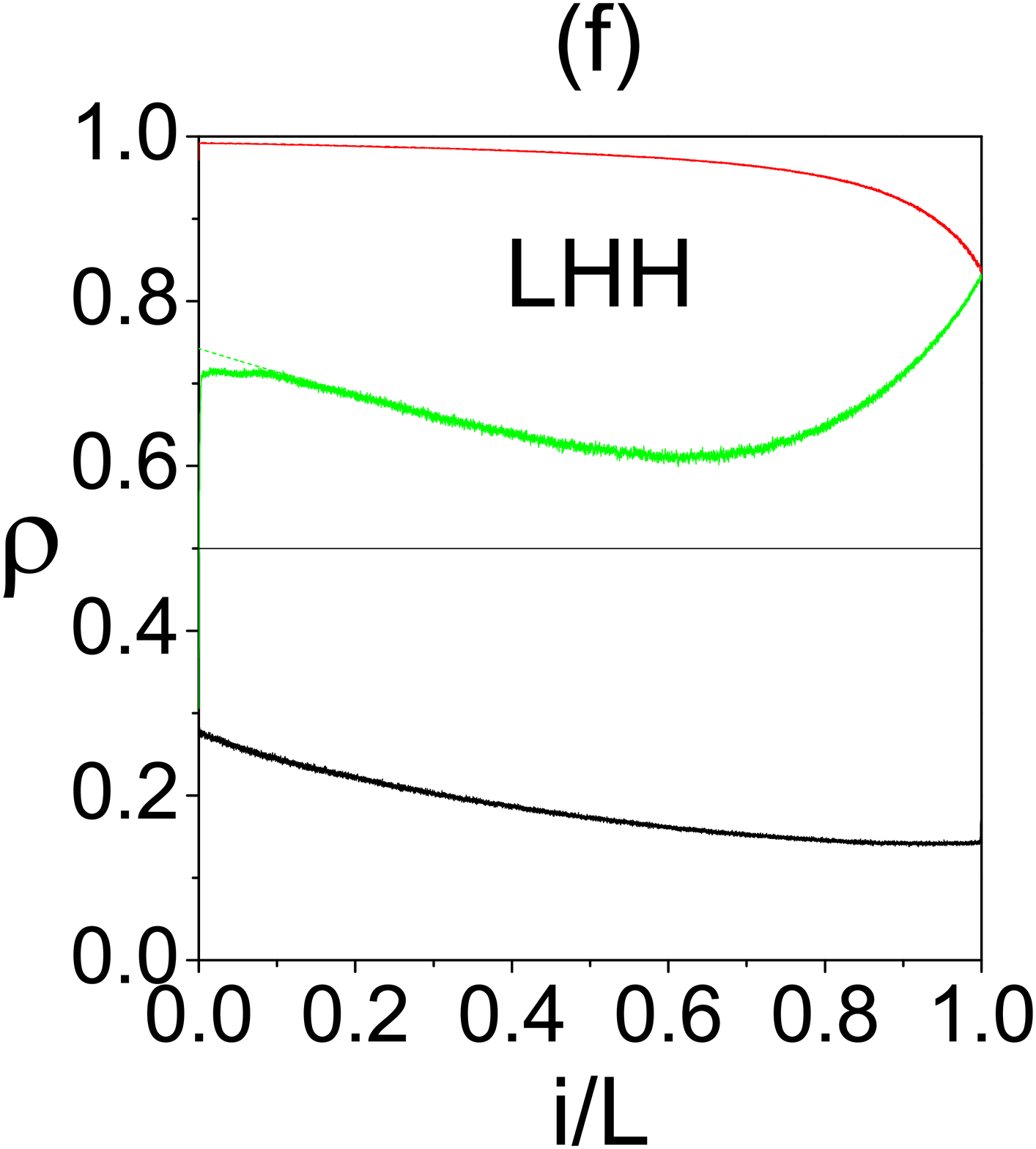}
\includegraphics[width=0.16\textwidth]{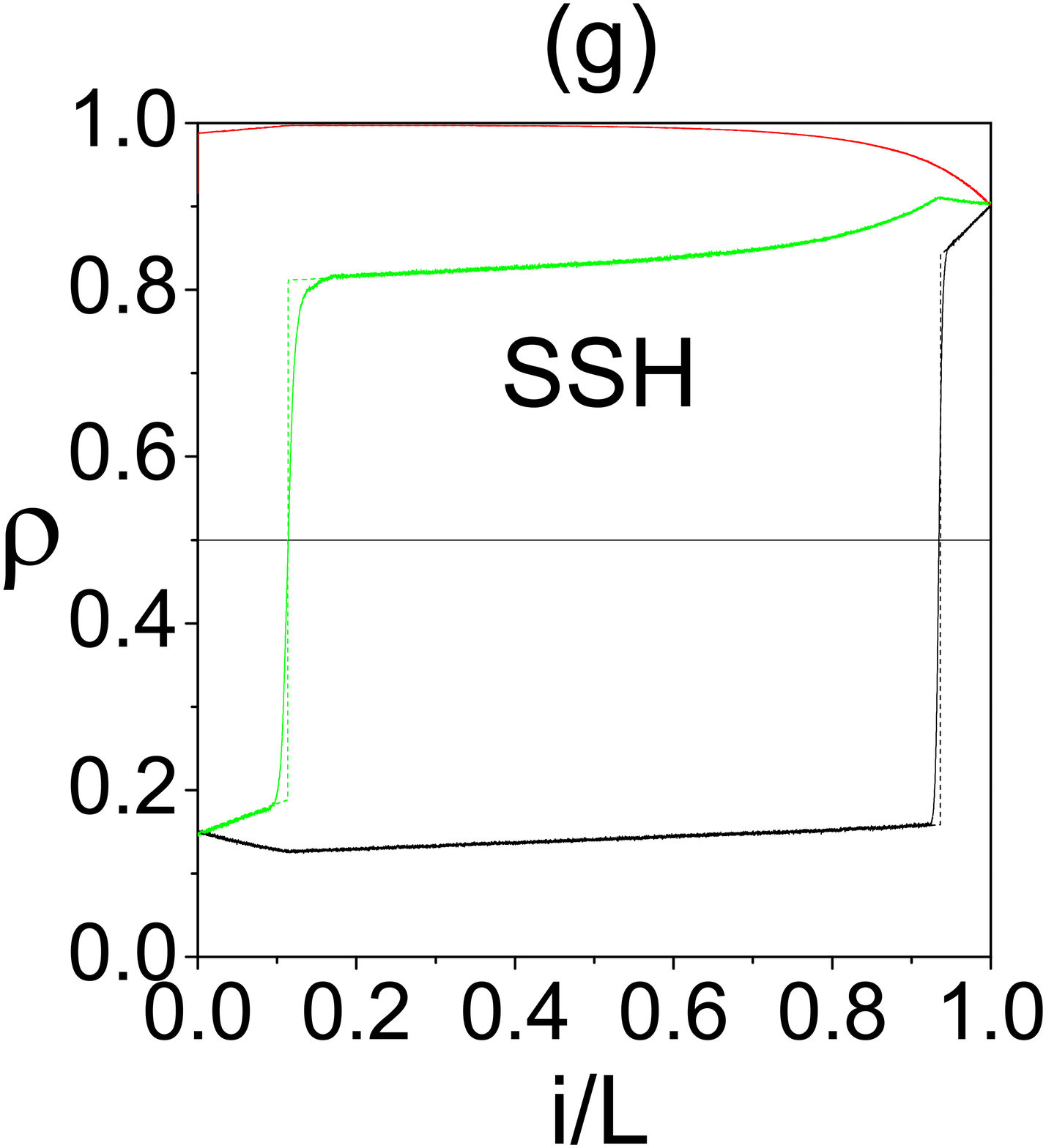}\includegraphics[width=0.16\textwidth]{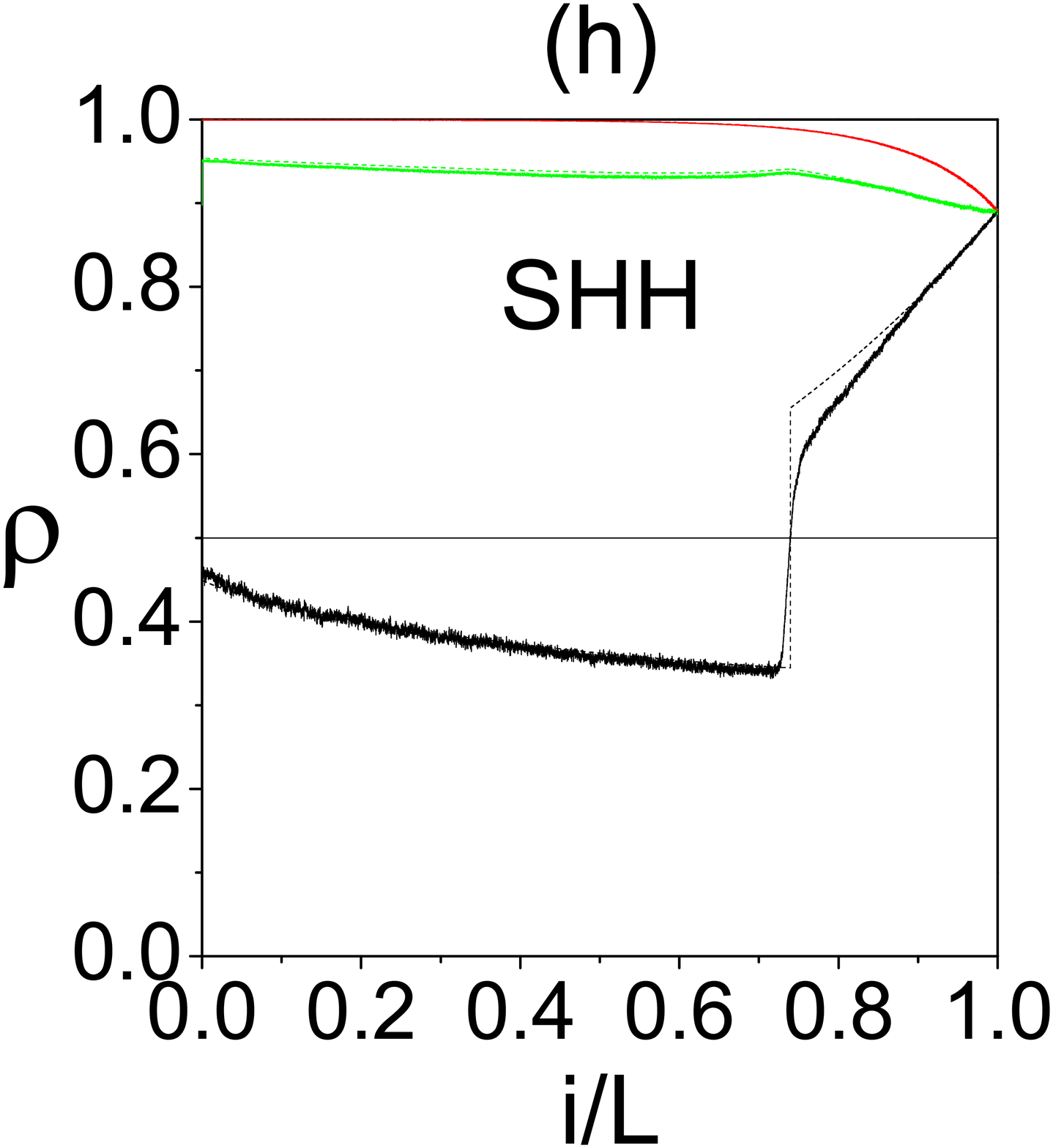}\includegraphics[width=0.16\textwidth]{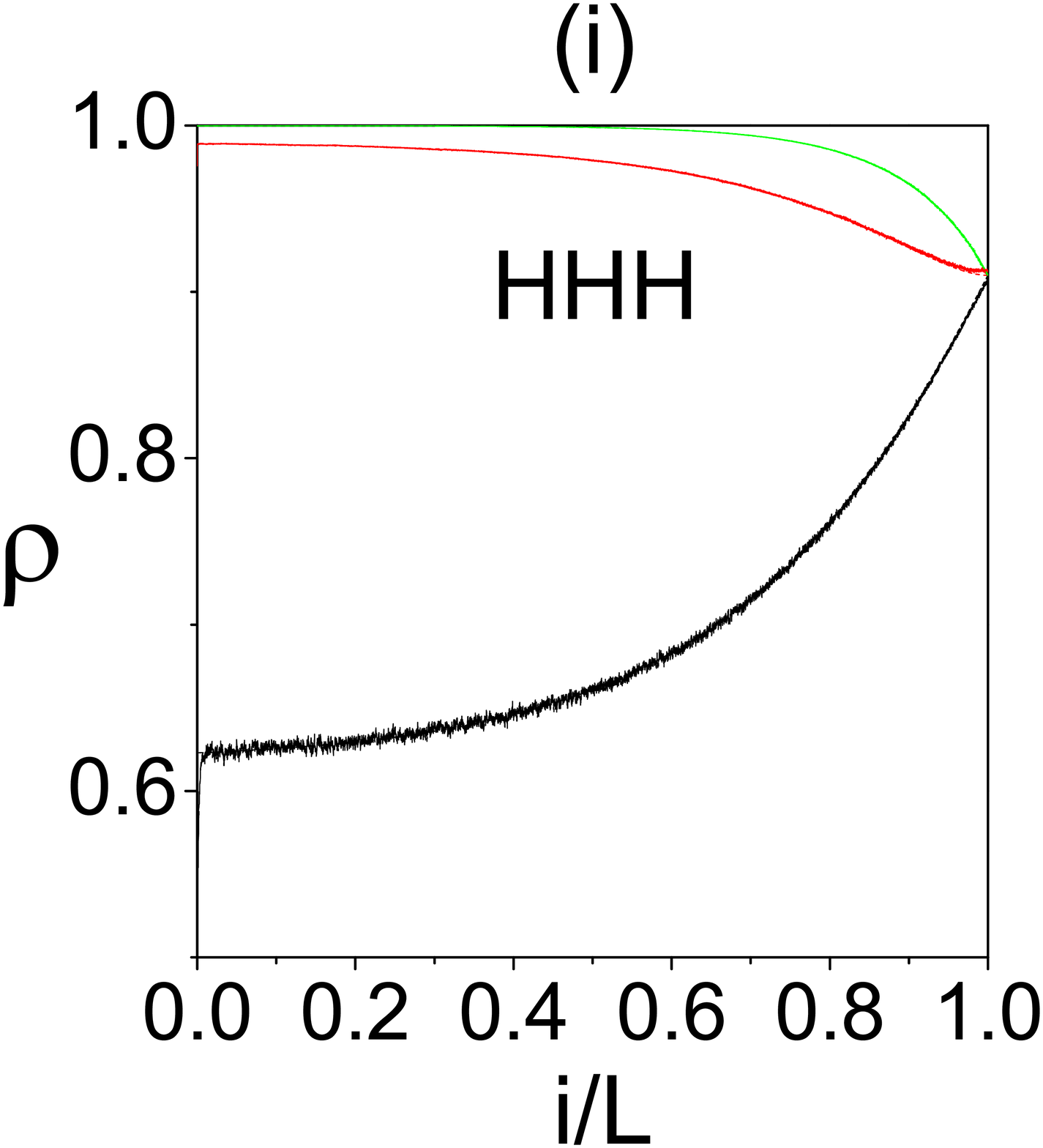}
\caption{(color online) Typical density profiles in state (a)
$LLL$, (b) $LLS$, (c) $LSS$, (d) $LLH$, (e) $LSH$, (f) $LHH$, (g)
$SSH$, (h) $SHH$, (i) $HHH$. The solid lines are simulation
results, and the dashed lines are mean field results. $(\alpha,
\beta)=$ (0.1, 0.25), (0.1, 0.15), (0.074, 0.087), (0.268, 0.286),
(0.2, 0.18), (0.278, 0.166), (0.15, 0.098), (0.45, 0.11), (0.45,
0.09) in (a)-(i), respectively. In the simulation, the system size
$N=20 000$.}
\end{figure}

\begin{figure}[!h]
\centering
\includegraphics[width=0.25\textwidth]{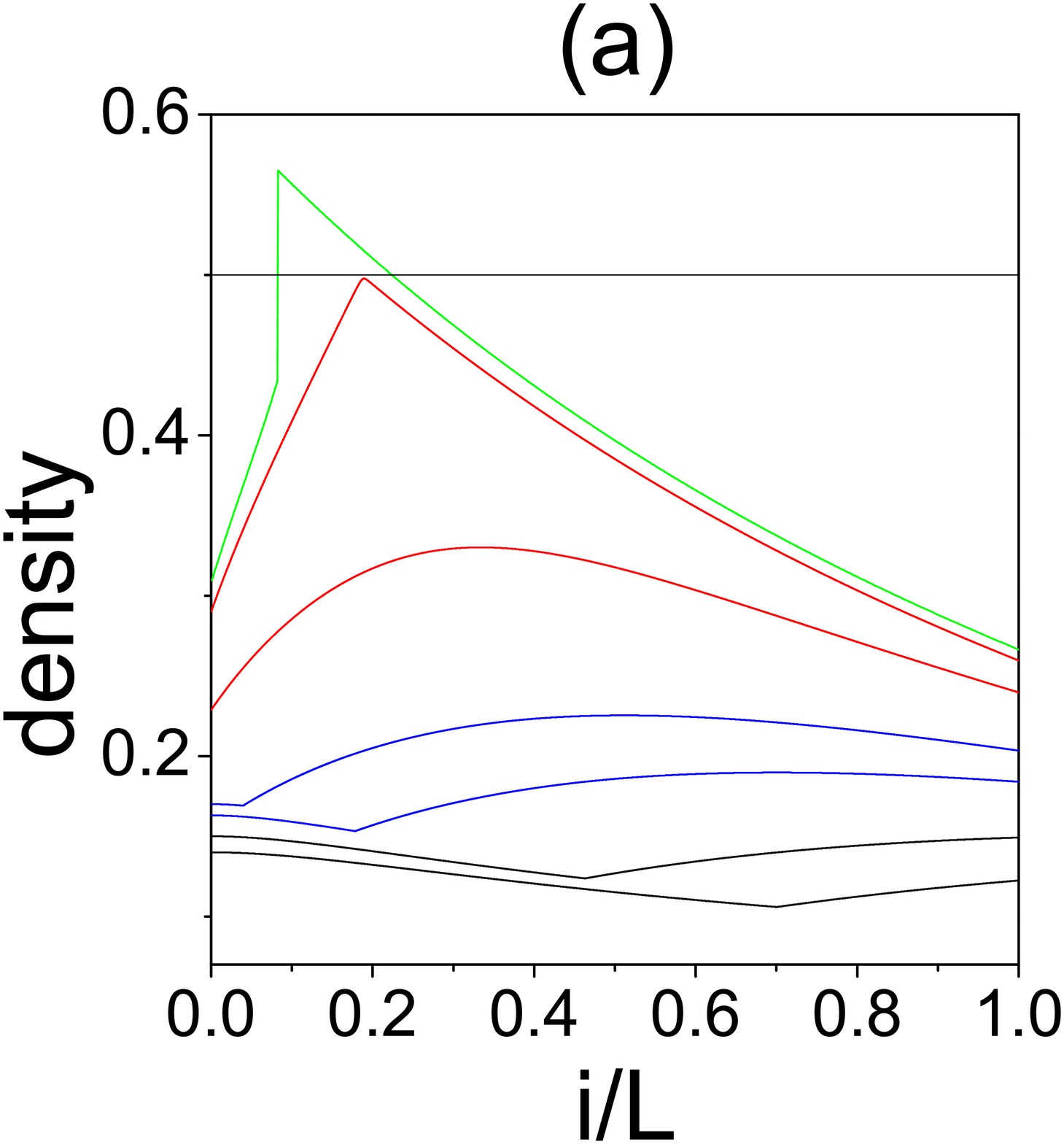}\hspace{-0.5cm}\includegraphics[width=0.25\textwidth]{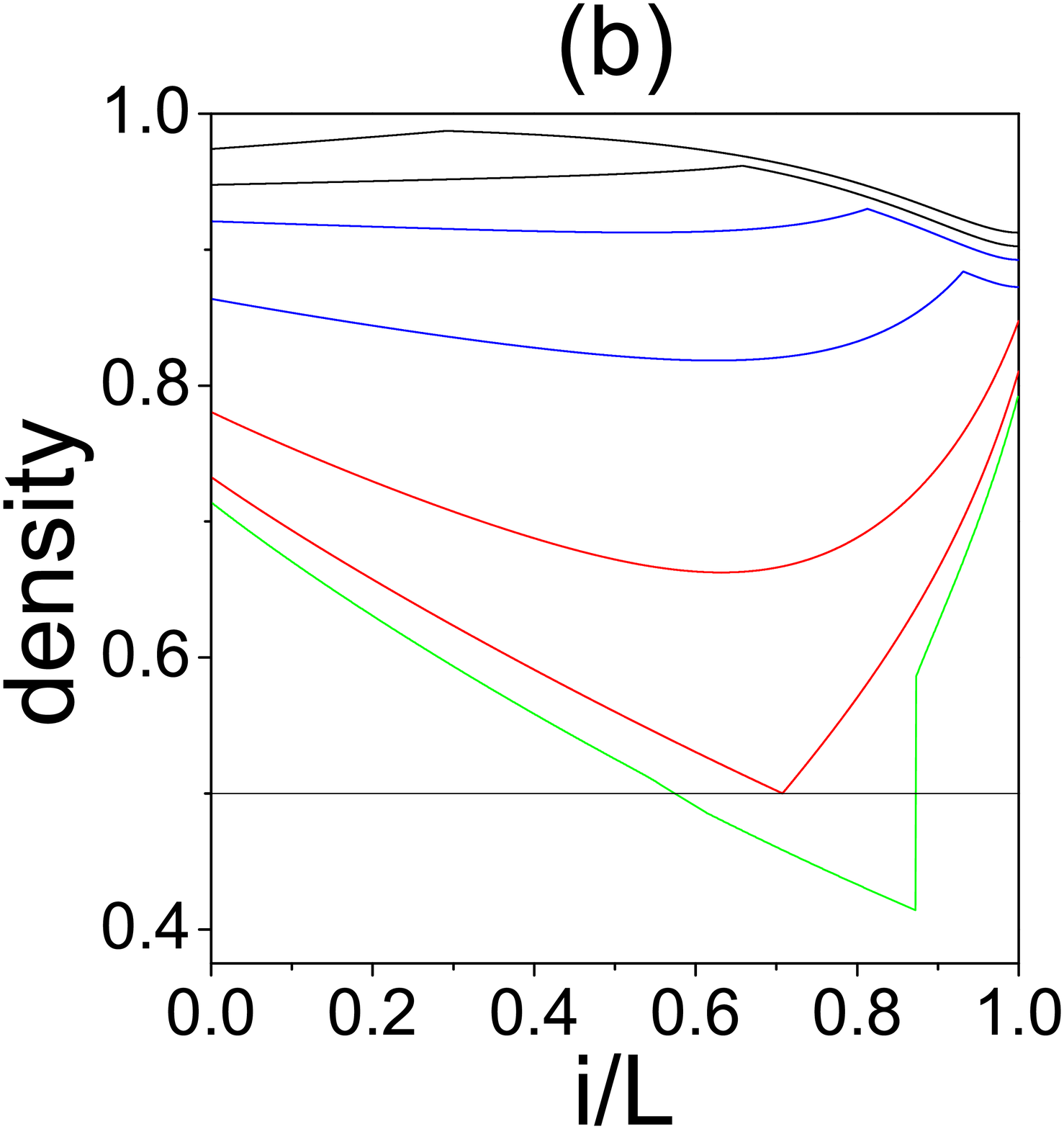}
\caption{ (color online) The density profiles of the middle lane
(mean field results). (a) System transits from LLS to $LS_1H$.
Parameters are $\beta=0.276$, $\alpha  = 0.14, 0.15, 0.163, 0.17,
0.229, 0.29, 0.308$ for curves from bottom to top. (b) System
transits from SHH to $LS_2H$. Parameters are $\alpha=0.310$,
$\beta  = 0.088, 0.098, 0.108, 0.128, 0.152, 0.189, 0.206$ for
curves from top to bottom. }
\end{figure}

We focus on the density profiles in the $LLS$ state. Fig.4(a)
shows several typical density profiles with the fixed value
$\beta=0.276$. Note that in the $LLS$ state, due to existence of a
shock on the third lane, the density profile is not smooth in the
middle lane at the shock location, which separates the density
profile into two parts. When $\alpha$ is left to the dotted line
in the phase diagram, the density profile in the downstream part
is increasing with $x$. With the increase of $\alpha$, the shock
moves left, so that the downstream part expands. Across the dotted
line, the density profile in the downstream part becomes
non-monotonically changing. This is because $J_{1\rightarrow
2}-J_{2\rightarrow 1}$ begins to change sign with $x$ in the
downstream part. Here $J_{i\rightarrow j}$ means flow from lane
$i$ to lane $j$. In the $LLH$ state, the shock has been expelled
out from the left end, and the density profile is always
monotonically changing. On the boundary between $LLH$ and $LS_1H$,
the extreme value of the density profile on lane 2 reached
$\rho=0.5$. As a result, a bulk induced phase transition occurs,
which exhibits a shock ($S_1$) as well as a CSDC density profile
downstream of the shock. Note that when $\rho_2=0.5$, the
coefficient $(1-2\rho_2)$ of Eq.(6) equals zero. Consequently, the
mean field equations (5)-(7) cannot be used to solve the CSDC
density profile in the vicinity of $\rho_2=0.5$, because the
numerical solution diverges. We have used Eqs.(2)-(4) instead.
Similarly, in the $SHH$ state, the density profile located in the
upstream part of the middle lane gradually becomes
non-monotonically changing when $(\alpha, \beta)$ approaches the
corresponding dotted line. Across the boundary between $LHH$ and
$LS_2H$, a bulk induced shock $(S_2)$ as well as a CSDC density
profile upstream of the shock appears, see Fig.4(b).

\begin{figure}[!h]
\centering
\includegraphics[width=0.25\textwidth]{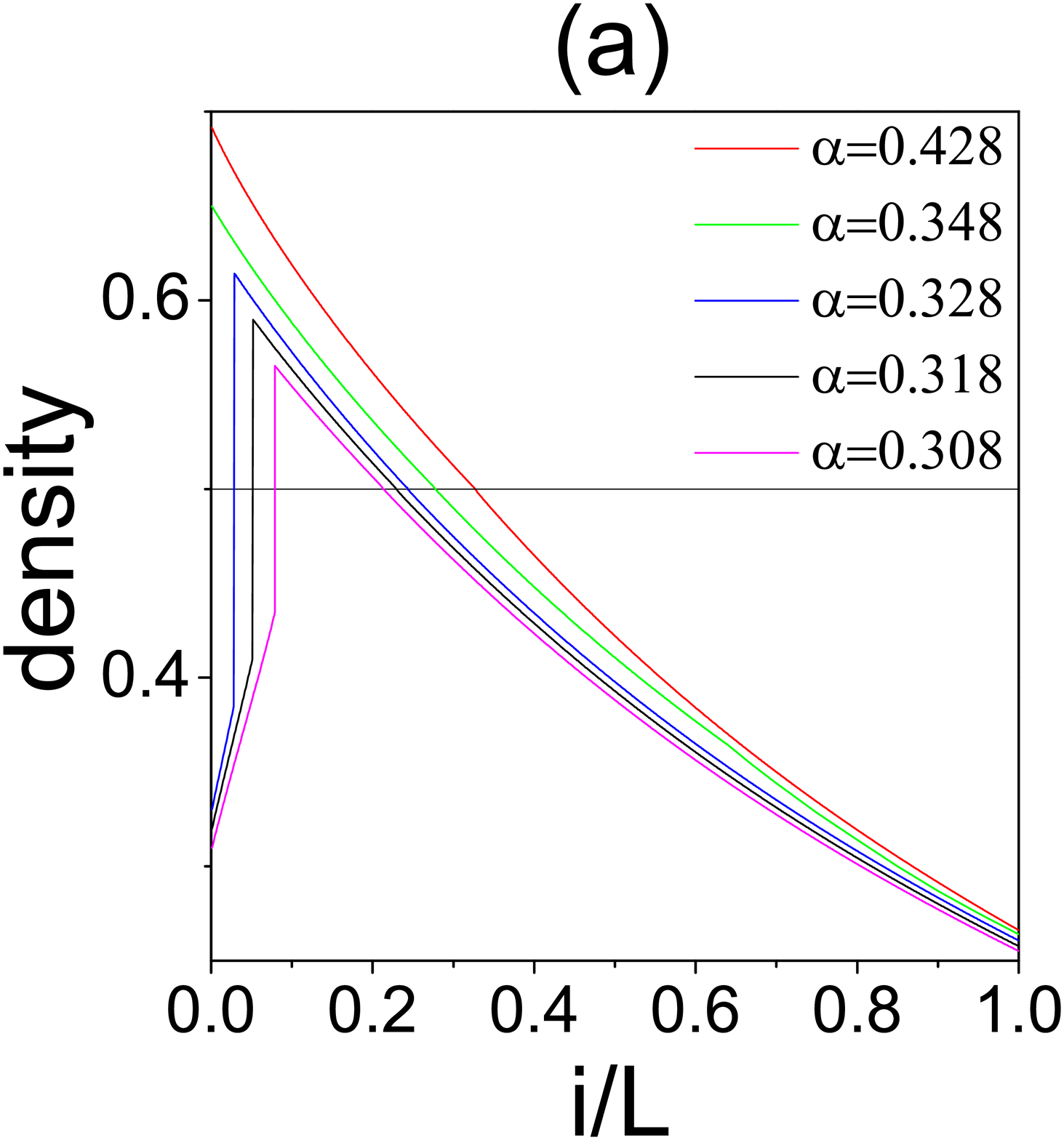}\hspace{-0.5cm}\includegraphics[width=0.25\textwidth]{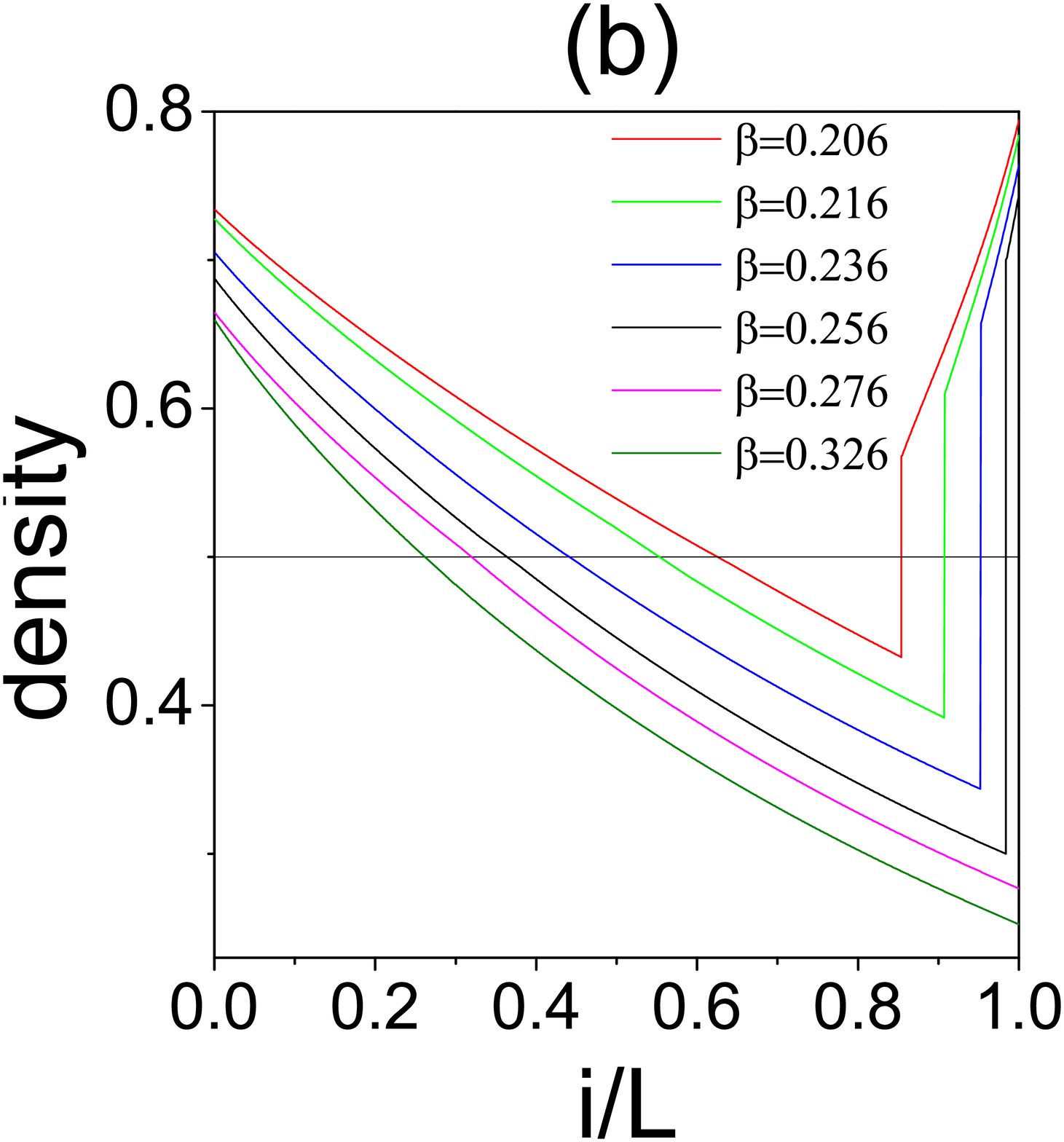}
\caption{(color online) The density profiles of the middle lane
(mean field results). (a) System transits from $LS_1H$ to $LCH$.
The parameter $\beta  = 0.296$. (b) System transits from $LS_2H$
to $LCH$. The parameter $\alpha = 0.358$. }
\end{figure}

\begin{figure}[!h]
\centering
\includegraphics[width=0.25\textwidth]{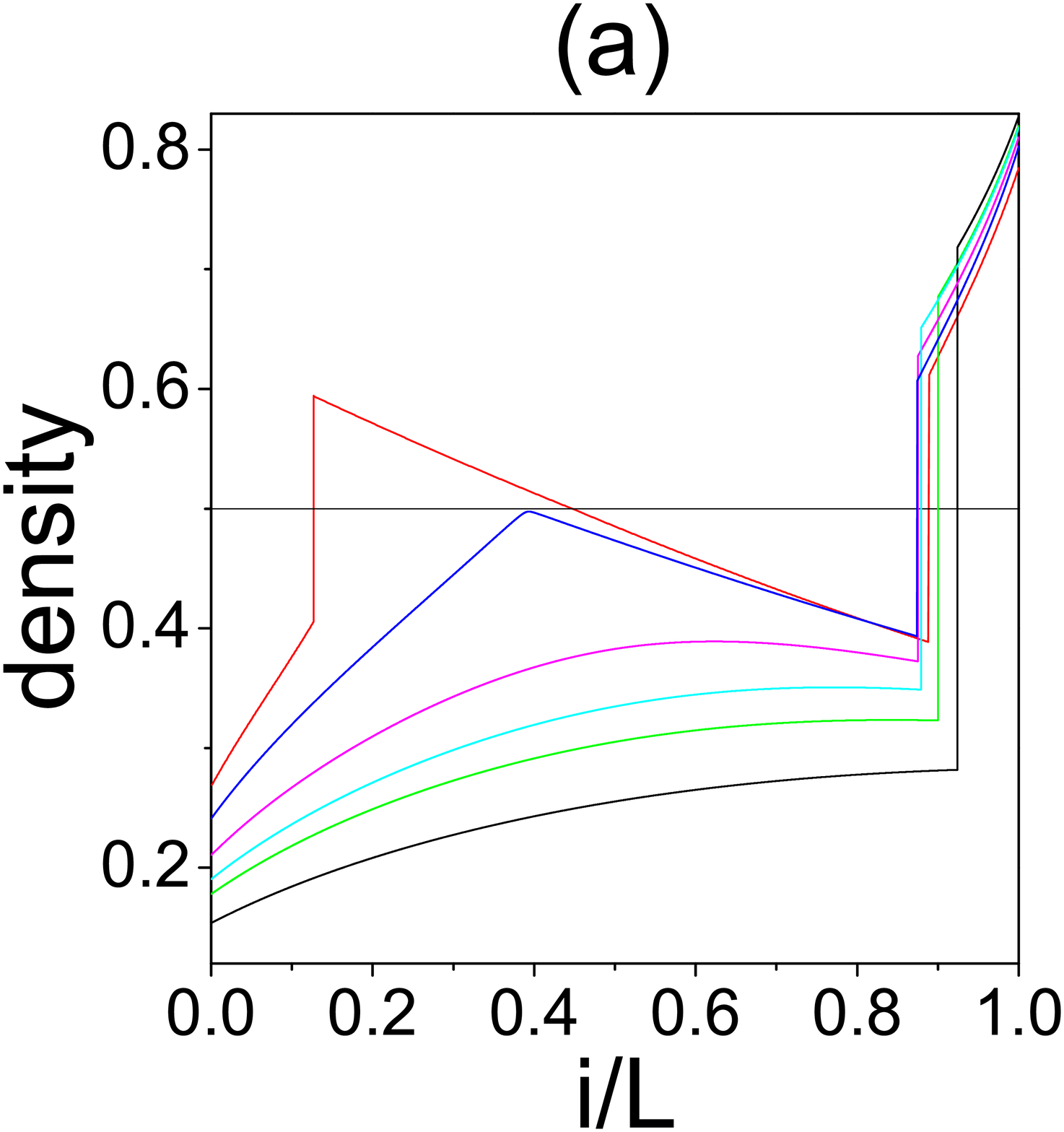}\hspace{-0.5cm}
\includegraphics[width=0.25\textwidth]{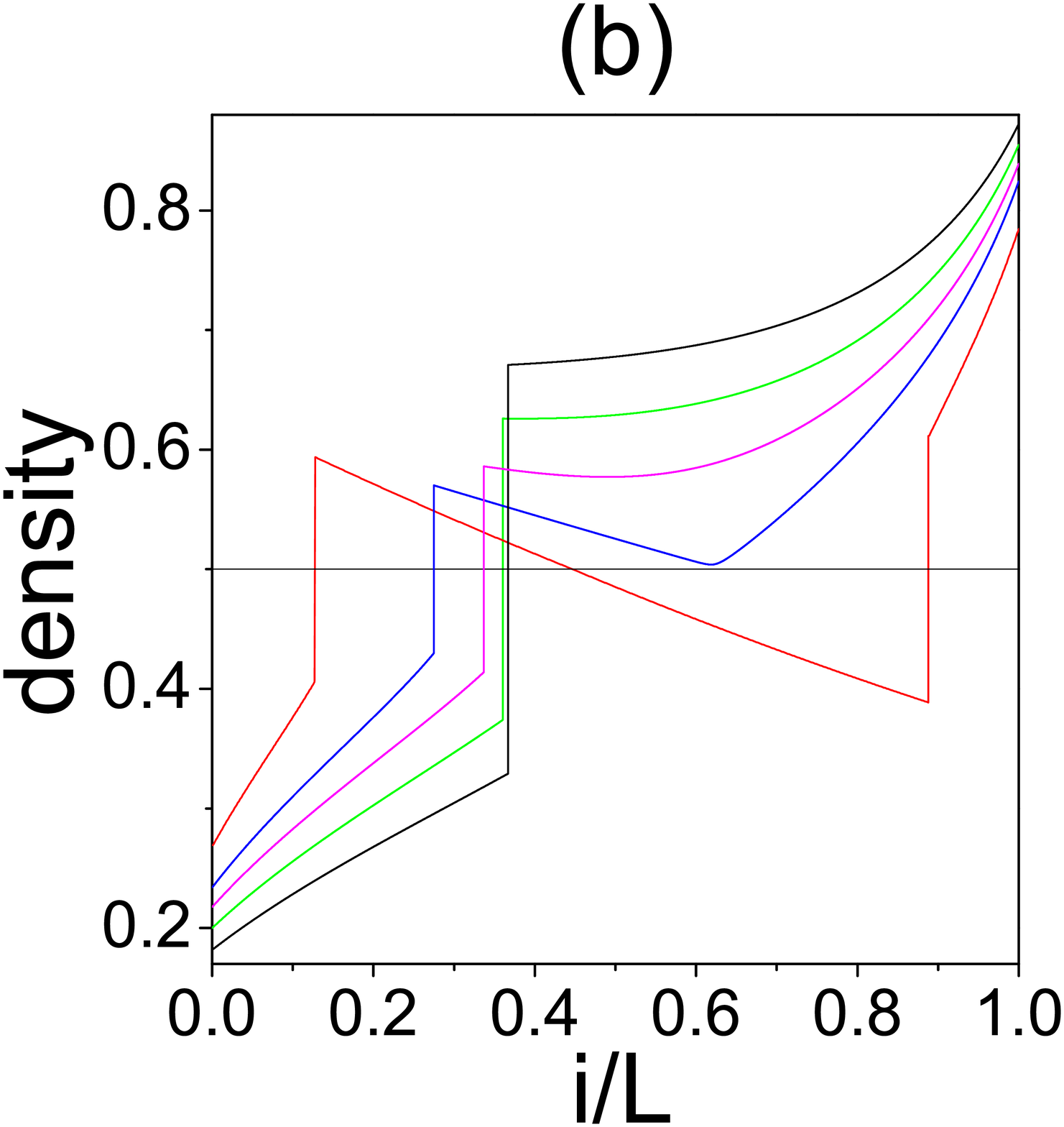}
\includegraphics[width=0.25\textwidth]{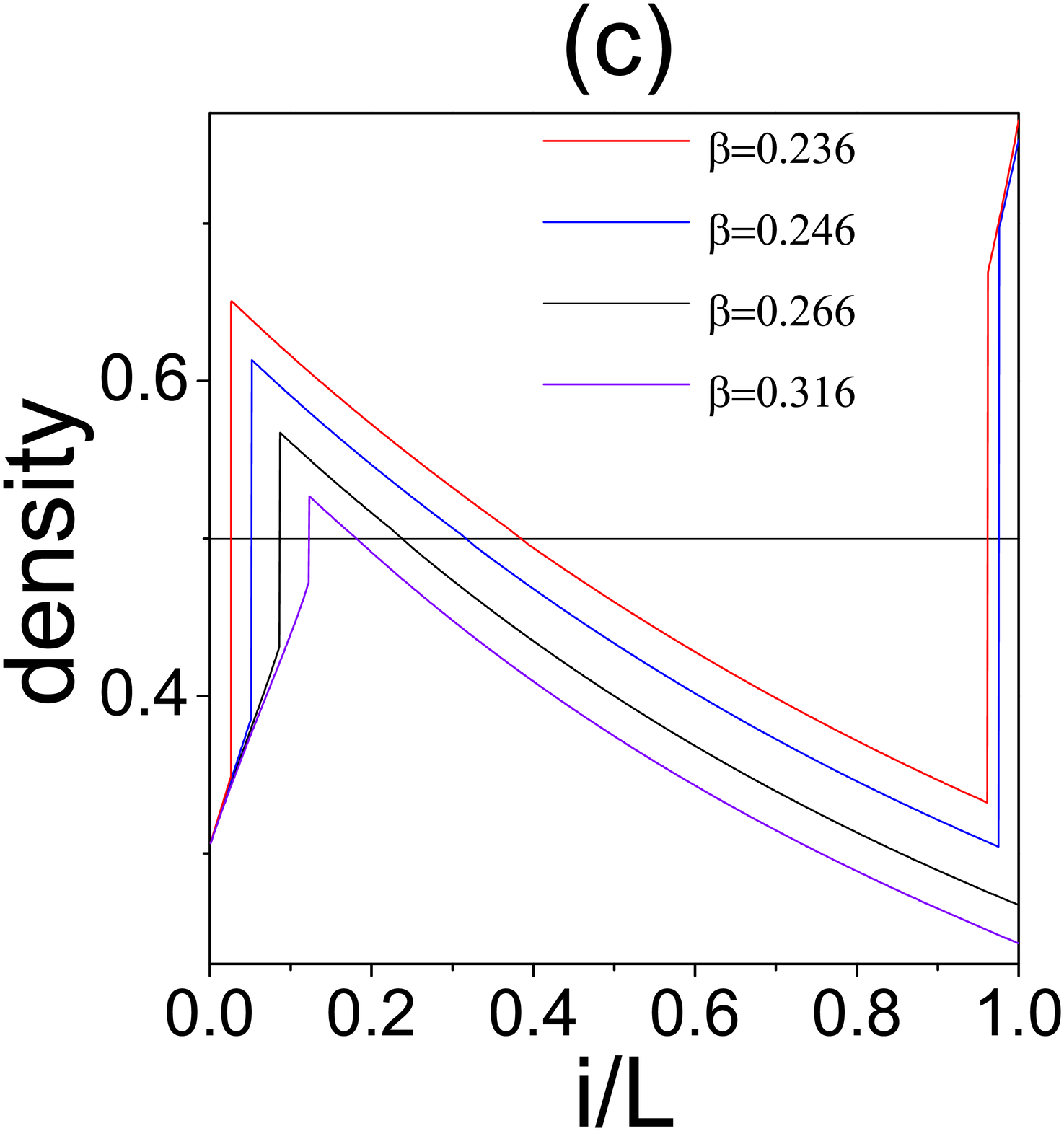}\hspace{-0.5cm}
\includegraphics[width=0.25\textwidth]{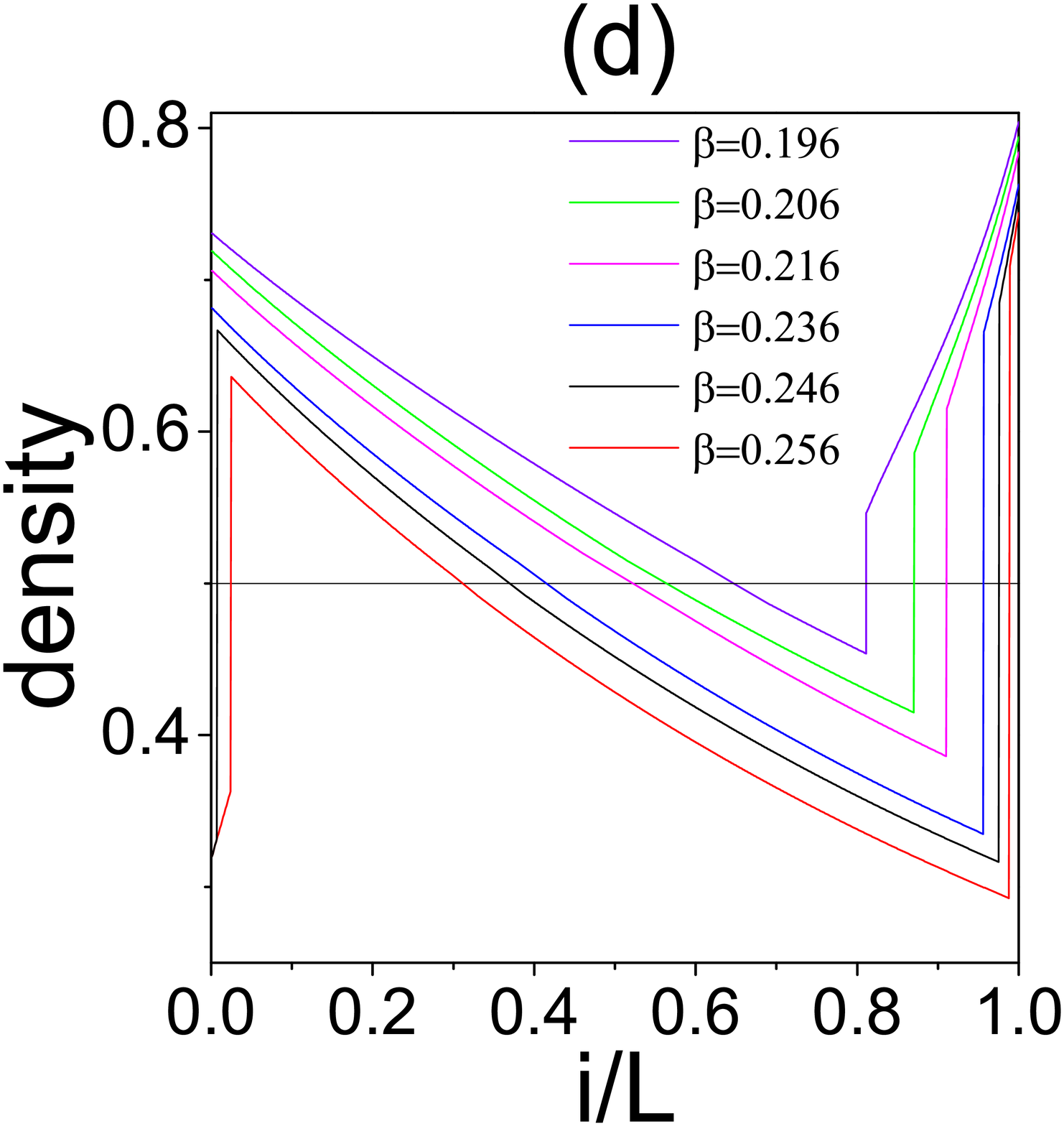}
\caption{(color online) The density profiles of the middle lane
(mean field results). (a) shows the transition from $LSH$ to $LDH$
along dashed line 1. Parameters are $(\alpha, \beta)=$ (0.154,
0.173), (0.178, 0.180), (0.190, 0.184), (0.211, 0.190), (0.241,
0.198) and (0.268, 0.206) for curves from bottom to top at $x=0$.
(b) shows the transition along dashed line 2. Parameters are
$(\alpha, \beta)=$ (0.182, 0.128), (0.200, 0.145), (0.218, 0.161),
(0.234, 0.175) and (0.268, 0.206) for curves from from top to
bottom at $x=1$.  (c) shows the transition from $LS_1H$ to $LDH$
along dashed line 3. The parameter $\alpha = 0.305$.  (d) shows
the transition from $LS_2H$ to $LDH$ along dashed line 4. The
parameter $\alpha = 0.318$. }
\end{figure}

\begin{figure}[!h]
\centering
\includegraphics[width=0.35\textwidth]{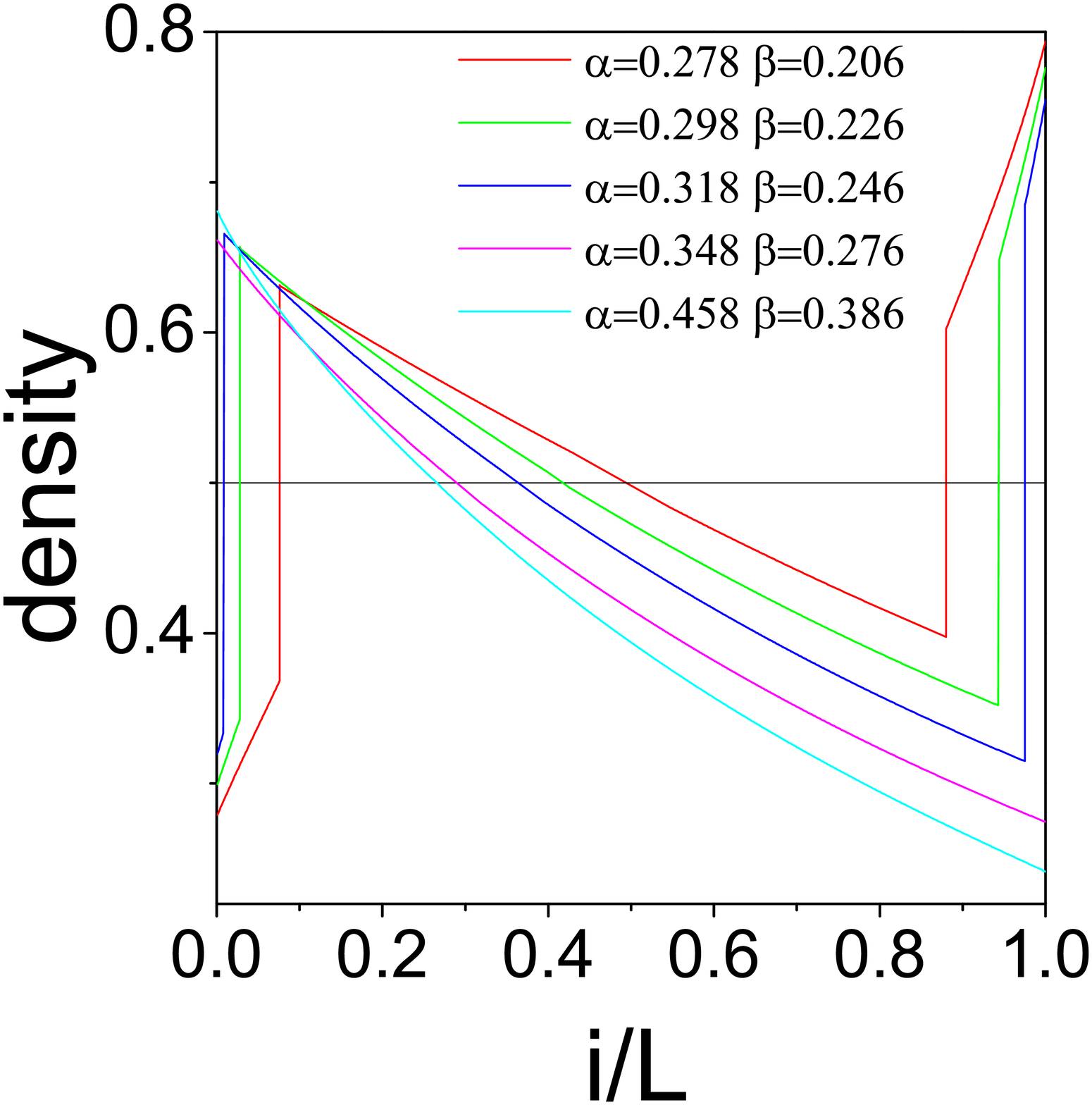}
\caption{ (color online) The density profiles of the middle lane
(mean field results). System transits from $LDH$ state to $LCH$
state along dashed line 5. }
\end{figure}

Next we interpret why only one bulk induced shock is triggered.
Suppose there are two bulk induced shock. As a result, the density
profile crosses $\rho=0.5$ twice, i.e., there are two locations
$x_1$ and $x_2$ ($x_1<x_2$) at which $\rho_2=0.5$. Therefore, we
have the following equation at the two locations according to the
current conservation principle
\begin{equation}
\rho_1(1-\rho_1)+\rho_3(1-\rho_3) = c,
\end{equation}
where $0< c\le 0.5$ stands for a constant.


Substituting $\rho_2=0.5$ into Eq.(6), we have
\begin{equation}
-\Omega_A(1-\rho_{3})+2\Omega_B\rho_{3}^2-\Omega_B(1-\rho_{1})+2\Omega_A\rho_{1}^2=0
\end{equation}
at locations $x_1$ and $x_2$.
Suppose $\Omega_A>\Omega_B$, then one has $\rho_1<0.5$ and
$\rho_3>0.5$ at the two locations. Thus, from Eq.(8), $\rho_3$ can
be solved
\begin{equation}
{\rho _3} = \frac{{1 + \sqrt {1 - 4[c - {\rho _1}(1 - {\rho _1})]} }}{2}
\end{equation}
Substituting Eq.(10) into (9), one has
\begin{equation}
\begin{split}
f\left( {{\rho _1}} \right) \equiv &- {\Omega _A}\left\{ {1 - \frac{{1 + \sqrt {1 - 4[c - {\rho _1}(1 - {\rho _1})]} }}{2}} \right\} \\
&+ 2{\Omega _B}{\left\{ {\frac{{1 + \sqrt {1 - 4[c - {\rho _1}(1 - {\rho _1})]} }}{2}} \right\}^2} \\
& - {\Omega _B}(1 - {\rho _1}) + 2{\Omega _A}\rho _1^2\\
&\left.{= 0}\right.\\
\end{split}
\end{equation}
Taking derivative to $\rho_1$, we obtain
\begin{equation}
\frac{{\partial f}}{{\partial {\rho _1}}} = \frac{{\left( {1 - 2{\rho _1}} \right)\left[ {{\Omega _A} + 2{\Omega _B}\left( {1 + d} \right)} \right]}}{d} + 4{\Omega _A}{\rho _1} + {\Omega _B}
\end{equation}
where $d = \sqrt {1 - 4\left[ {c - {\rho _1}\left( {1 - {\rho _1}}
\right)} \right]}$. Since $\rho_1<0.5$, $\partial f/\partial
\rho_1>0$ is always satisfied. This means that Eq.(11) has at most
one solution. Therefore, $\rho_1(x_1)=\rho_1(x_2)$ and
$\rho_3(x_1)=\rho_3(x_2)$. When one lane-changing parameter equals
0, both $\rho_1(x)$ and $\rho_3(x)$ are monotonically changing
with $x$. Therefore, $\rho_1(x_1)=\rho_1(x_2)$ and
$\rho_3(x_1)=\rho_3(x_2)$ cannot be satisfied.

In the general situation that neither $\Omega_A$ nor $\Omega_B$
equal to 0, the section $x\le x_1$ and the section $x\ge x_2$ can
match each other and constitute a shortened system with length
$1-(x_2-x_1)$, in which only one bulk induced shock exists. The
mean-field analysis has shown that given the same value of
$\alpha$ and $\beta$, the flow rate in a shortened system is
larger than that in the original system, provided there exists one
bulk induced shock in both systems. Thus, we argue that the
current minimization principle excludes the occurrence of two or
more bulk induced shocks.

We study the density profiles in the $LS_1H$ state. Fig.5(a) shows
several density profiles with $\beta=0.296$ is fixed. With the
increase of $\alpha$, the shock gradually moves left. Across the
boundary between $LS_1H$ and $LCH$, the shock is expelled out from
the left end, and only the CSDC density profile is left in the
middle lane. Similarly, the shock is expelled out from the right
end in the $LS_2H$ phase when across the boundary between $LS_2H$
and $LCH$, see Fig.5(b).

Figs.6 (a) and (b) show the density profiles in the $LSH$ phase,
in which $(\alpha, \beta)$  changes along dashed lines 1 and 2 in
the phase diagram, respectively. When $(\alpha, \beta)$ is left of
the dotted line, the density profile is increasing both upstream
and downstream of the shock. However, when across the dotted line,
the density profile becomes non-monotonic upstream (downstream) of
the shock. On the boundary between $LSH$ and $LDH$, the maximum
(minimum) of density profile reaches $\rho=0.5$. When across the
boundary, the bulk induced shock appears upstream (downstream) of
the first shock and thus two shocks exist simultaneously in the
middle lane.

Figs.6(c) and (d) show the density profiles in the $LS_1H$ and
$LS_2H$ phase, in which $(\alpha, \beta)$  changes along dashed
lines 3 and 4, respectively. On approaching the boundary between
$LS_1H$ ($LS_2H$) and $LDH$ , $\rho_2(1)$ ($\rho_2(0)$) gradually
approaches $\beta$ (($1-\alpha$)). Across the boundary, a shock is
induced from the right (left) boundary, thus, double shocks
emerge.

Finally, in the $LDH$ phase, when $(\alpha, \beta)$  changes along
dashed line 5, the left shock moves toward left and the right
shock moves right. On the boundary between the $LDH$ phase and the
$LCH$ phase, the two shocks are expelled from the system,
simultaneously, see Fig.7.

{\it Discussion.} Hinsch and Frey have studied a periodic
one-dimensional exclusion process composed of a driven and a
diffusive part, and identified bulk-driven phase transitions in a
mesoscopic limit where both dynamics compete [30]. Nevertheless,
the system can be regarded as two sub-systems connecting together.
Therefore, the bulk-driven phase transitions are essentially
boundary induced ones.

In our system, we can also treat the location where the CSDC
density profile crosses $\rho=0.5$ as a virtual boundary, which
separates the system into two sub-systems. For the left
sub-system, the effective exit rate for the middle lane is
$\beta_{2,eff}=0.5$. For the right one, the effective entrance
rate for the middle lane is $\alpha_{2,eff}=0.5$. Nevertheless,
instead of a static boundary, the location of the virtual boundary
in our system is self-tuned and determined by the values of the
kinetic rates.

We also would like to point out that the existence of double
shocks has been demonstrated when considering detachment and
attachment of particles in the Katz-Lebowitz-Spohn process [31].
However, different from the double shocks in $LDH$ state in our
model, both shocks are boundary induced ones in Ref.[31].

{\it Conclusion.} To summarize, we have studied a weakly and
asymmetrically coupled three-lane TASEP. The phase diagram has
been presented. A non-monotonically changing density profile in
the middle lane has been observed in the $LLH$, $LHH$, $LSH$
phases. When the extreme value of the density profile reaches
$\rho=0.5$, a bulk induced phase transition occurs which exhibits
a shock and a CSDC density profile upstream or downstream of the
shock. In particular, the CSDC density profile has not been
observed before. The double shocks comprising of one bulk induced
shock and one boundary induced shock has also been observed. In
the situation that one of the lane changing rates equals zero, it
can be easily proved that there cannot exist two bulk induced
shocks. In the general case where both lane changing rates are
nonzeros, the current minimization principle has excluded the
occurrence of two or more bulk induced shocks.

This paper only focuses on the three-lane TASEP. When the number
of lanes further increases, the phase diagram will become much
more complicated. The questions such as whether two or more bulk
induced shocks could be observed simultaneously in a lane, whether
bulk induced shocks can be observed simultaneously in different
lanes, need to be investigated in the future work.

\section*{ACKNOWLEDGMENTS}
This work is funded by the National Basic Research Program of
China (No.2012CB725404), the National Natural Science Foundation
of China (Grant Nos. 71371175 and 71171185).


\begin{thebibliography}{11}

\bibitem{1}
B. Schmittmann, R.K.P. Zia, Phys. Rep. {\bf 301}, 45 (1998).

\bibitem{2}
 B. Derrida, Phys. Rep. \textbf{301}, 65 (1998).

 \bibitem{3}
 G. M. Sch\"{u}tz,  in \textit{Phase Transitions and Critical Phenomena}, edited by C. Domb and J.  Lebowitz
(Academic, London, 2000), vol. 19.

\bibitem{4}
 R. A. Blythe,  M. R. Evans, J. Phys. A \textbf{46}, R333 (2007)

\bibitem{5}

T. Chou, K. Mallick and R. K. P. Zia, Rep. Prog. Phys. {\bf 74},
116601 (2011)



\bibitem{8}
M. R. Evans, D. P. Foster, C. Godreche and D. Mukamel, Phys. Rev.
Lett. {\bf 74}, 208 (1995); J. Stat. Phys. {\bf 80}, 69 (1995).

\bibitem{9}
 M. R. Evans, Y. Kafri,  H. M. Koduvely, and D. Mukamel, Phys. Rev. Lett. {\bf 80}, 425 (1998).

\bibitem{10}
Y. Kafri, E. Levine, D. Mukamel,
 Phys. Rev. Lett. {\bf 89}, 035702 (2002).

\bibitem{9a}

 Arndt, P.F., Heinzel, T., Rittenberg, V.: J. Stat. Phys. 90, 783
 (1998); Rajewsky, N., Sasamoto, T., Speer, E.R.: Physica A 279, 123 (2000)

 \bibitem{10a}

Adams, D.A., Schmittmann, B., Zia, R.K.P.: Phys. Rev. E 75, 041123
(2007); Korniss, G., Schmittmann, B., Zia, R.K.P.: Europhys. Lett.
45, 431 (1999); Mettetal, J.T., Schmittmann, B., Zia, R.K.P.:
Europhys. Lett. 58, 653 (2002)



\bibitem{6}
J. Krug, Phys. Rev. Lett. {\bf 67}, 1882 (1991).

\bibitem{7}
V. Popkov and G. M. Sch\"{u}tz, Europhys. Lett. 48, 257 (1999).

\bibitem{11}
A. B. Kolomeisky, G. M. Sch\"{u}tz, E. B. Kolomeisky, and J. P.
Straley, J. Phys. A 31, 6911 (1998)


\bibitem{12}
J. S. Hager, J. Krug, V. Popkov, and G. M. Sch\"{u}tz. Phys. Rev.
E \textbf{63}, 056110 (2001).



\bibitem{12a}
A. Parmeggiani, T. Franosch, E. Frey,  Phys. Rev. Lett.
\textbf{90}, 086601 (2003); Phys. Rev. E 70 046101 (2004).




\bibitem{13}
 M. R. Evans, Y. Kafri, K. E. P. Sugden, J. Tailleur, J. Stat. Mech. P06009 (2011).


\bibitem{13a}
 R. J. Harris, R. B. Stinchcombe,  Physica A \textbf{354}, 582 (2005).

\bibitem{14}
T. Reichenbach,  T. Franosch, E. Frey, Phys. Rev. Lett.
\textbf{97}, 050603 (2006).

\bibitem{15}
R. Juh\'{a}sz,   Phys. Rev. E \textbf{76}, 021117 (2007); J. Stat.
Mech. P03010 (2010).

\bibitem{16}
 R. Jiang, M. B. Hu, Y. H. Wu, Q. S. Wu,  Phys. Rev. E \textbf{77}, 041128 (2008).

\bibitem{17}
R. Jiang, K. Nishinari, M. B. Hu, Y. H. Wu, and Q. S. Wu,  J.
Stat. Phys. \textbf{136}, 73 (2009).

\bibitem{18}
Z. P. Cai, Y. M. Yuan, R. Jiang et al.,  J. Stat. Mech. P07016
(2008).

\bibitem{19}
 K. Tsekouras,  A. B. Kolomeisky,  J. Phys. A \textbf{41},
095002 (2008); \textbf{41}, 465001 (2008); E. Pronina and A. B.
Kolomeisky, J. Phys. A {\bf 37}, 9907 (2004).


\bibitem{20}
V. Popkov, I. Peschel,  Phys. Rev. E \textbf{64}, 026126 (2001);
V.Popkov, M.Salerno,  Phys. Rev. E \textbf{83}, 011130 (2011).

\bibitem{21}
I. T. Georgiev,  B. Schmittmann, R. K. P. Zia,   Phys. Rev. Lett.
\textbf{94}, 115701 (2005).


\bibitem{22}
A. Melbinger, T. Reichenbach, T. Franosch, and E. Frey, Phys. Rev.
E \textbf{83}, 031923 (2011).


\bibitem{23}
C. Schikmann, C. Appert-Rolland, and L. Santen,  J. Stat. Mech.
P06002 (2010).


\bibitem{24}
Q. H. Shi, R. Jiang, M. B. Hu, Q. S. Wu, J.Stat.Phys. {\bf 142},
616 (2011).

\bibitem{25}
T. Ezaki and K. Nishinari,  Phys. Rev. E \textbf{84}, 061141
(2011).


\bibitem{26}
H. Hinsch and E. Frey,  Phys. Rev. Lett. \textbf{97}, 095701
(2006).


\bibitem{28}
 V. Popkov, A. Rakos, R. D. Willmann, A. B.
Kolomeisky, and G. Sch\"{u}tz, Phys. Rev. E 67, 066117 (2003).

\end{thebibliography}
\end{document}